\documentclass[11pt]{article}
\usepackage{amsmath}
\usepackage{amssymb}
\usepackage{epsfig}
\usepackage{verbatim}
\begin{document}
\title{{\bf{\Large Rainbow Rindler metric and Unruh effect}}}
\author{ 
{\bf {\normalsize Gaurav Yadav}$
$\thanks{E-mail: gaurav.yadav@iitg.ernet.in}},\,\,\
{\bf {\normalsize Baby Komal}$
$\thanks{E-mail: baby.komal@iitg.ernet.in}}\,\,\ and 
{\bf {\normalsize Bibhas Ranjan Majhi}$
$\thanks{E-mail: bibhas.majhi@iitg.ernet.in}}\\ 
{\normalsize Department of Physics, Indian Institute of Technology Guwahati,}
\\{\normalsize Guwahati 781039, Assam, India}
\\[0.3cm]
}
\maketitle

\begin{abstract}
The energy of a particle moving on a spacetime, in principle, can affect the background metric. The modifications to it depend on the ratio of energy of the particle and the Planck energy, known as rainbow gravity. Here we find the explicit expressions for the coordinate transformations from rainbow Minkowski spacetime to accelerated frame. The corresponding metric is also obtained which we call as rainbow-Rindler metric. So far we are aware of, no body has done it in a concrete manner. Here this is found from the first principle and hence all the parameters are properly identified. The advantage of this is that the calculated Unruh temperature is compatible with the Hawking temperature of the rainbow black hole horizon, obtained earlier. Since the accelerated frame has several importance in revealing various properties of gravity, we believe that the present result will not only fill that gap, but also help to explore different aspects of rainbow gravity paradigm.   
\end{abstract}
\vskip 9mm

\section{Introduction}
The nature of reality at the Planck scale is the subject of much debate in the recent world of physics. For energies approaching the Planck scale energy, a new class of phenomena are expected and we require a better theory for spacetime that can explain these phenomena as the classical descriptions are no more applicable. The different approaches to attack this issue are like loop quantum gravity \cite{1,2}, string theory \cite{3,4}, Lorentzian dynamical triangulations \cite{51}-\cite{55}, non commutative geometry \cite{6}-\cite{65}, condensed matter analogues \cite{7}-\cite{74}, etc. To workout the various quantities at the Planck scale we have taken the approach as provided by Double Special Relativity \cite{8,9,92,10,101} wherein a moving particle deforms the space-time such that the dispersion relation becomes energy dependent in the following way,
\begin{equation}
 E^2\big(f(E/E_p)\big)^2 - c^2p^2\big(g(E/E_p)\big)^2 = m^2c^4
\label{mdr} 
 \end{equation}
The above modification leads to a situation where we have an observer-independent length scale in addition to a velocity scale, in the Planck scale regime. This generalization of the Minkowski space implies to a further generalized Lorentz transformation which reduces to the old form under suitable limits. The deformed Lorentz transformations have already been given in \cite{16}, in terms of Lorentz boosted energy where the spacetime Lorentz transformations depend upon the energy. Also, in the low energy limit $(E/E_p \ll 1)$, the metric coefficients reduce to normal Minkowski metric. However, rainbow Rindler transformations has not been properly dealt with.  

   The one of the importances of Rindler metric lies in black hole thermodynamics. Equivalence principle provides us a way to explore several general relativistic (GR) features by going into the accelerated frame. Due to simplicity of the structure of the metric, it is always useful in doing exact analytic calculations which are sometimes very hard or impossible in real curved spacetimes. Moreover, the black hole spacetimes reduce to Rindler form in the near horizon limit. Thus such a metric demands a special attention in gravitational theory. Recently, the rainbow gravity has attracted a lot of attention and people are looking at it in different angles \cite{18}--\cite{Ali:2014yea}. Particularly, the modifications to thermodynamic quantities of the black holes and its implications draw attention. (For more on rainbow effect on different areas of gravity, see \cite{Kim:2016qtp}--\cite{Leiva:2008fd}.) This immediately brings in our mind to find the form of the Rindler metric in the context of rainbow gravity. In literature, there are some sporadic attempts \cite{Ling:2006ba}, but none of them are complete or very precise in terms of proper identification of precise parameters, like acceleration etc. The transformations were given in a loose manner. The reason behind this is, no one has derived it from the first principle. In this paper we shall fill this gap.       
    
    Here, we adopt a framework where we will explicitly verify the Lorentz transformations and also give a detailed analysis of deformed spacetime in Rindler frame. Rainbow Rindler transformations will be obtained by evaluating the four momentum of the moving frame with respect to the inertial observer. We will also give an explicit form of the effective acceleration in terms of usual acceleration (i.e. in absence of rainbow parameters) of moving observer. Next, using them the Unruh temperature will be calculated by different approaches. It will be observed that our finding is compatible with the previously derived horizon temperature of a black hole in rainbow gravity \cite{18,Gim:2014ira}. Thus the analysis verifies the correctness of our calculations and results.
    
  The organization of the paper is as follows. In section \ref{Setup}, we shall give the functional forms of the transformations from rainbow Minkowski spacetime to any moving frame. In the next section, we will use them to verify the already existing Lorentz transformations. Section \ref{Rindler} will be our main part of this paper. Here the explicit forms of Rindler transformations and corresponding accelerated metric will be derived in the context of Rainbow paradigm. Then the Unruh effect will be discussed in section \ref{Unruh} in three different subsections corresponding to three alternative approaches. Here we shall give only the relevant steps for convenience of the readers since the calculation is almost similar to the usual one. Final section will be devoted for the conclusions.

Notation: In this paper we shall adopt the following notations. The Latin indices $a,b$, etc. denote all the space-time coordinates while Greek indices $\alpha,\beta$, etc. refer to space coordinates only.

\section{\label{Setup}General setup}
The simplest generalization of the space-time metric $ g_{ab}$ which corresponds to the modified dispersion relation (\ref{mdr}), under the assumption of an isotropic space \cite{16}, is given by,
\begin{equation}
g_{ab}= diag~(-1/f^2,1/g^2,1/g^2,1/g^2)
\end{equation}   
Given the constancy of the speed of light, the most natural procedure will be to use light signals to set up the coordinates. 
 To stress this fact and with future applications in mind, we will first obtain the coordinate transformations for the general observer and then specialize to an observer moving with a uniform velocity.
 
    The rainbow metric for Minkowski spacetime is given by \cite{12},
\begin{eqnarray}
ds^2 &=& g_{ab}dx^a dx^b = -\frac{{c^2}{dT^2}}{f^2} + \frac{dX^2}{g^2} + \frac{dy^2}{g^2}+\frac{dz^2}{g^2}
\nonumber
\\ 
&=& -\frac{{c^2}{dT^2}}{f^2} + g_{\alpha\beta}dx^\alpha dx^\beta~,
\label{1.01}
\end{eqnarray}
where $\text{diag}(g_{\alpha\beta}) = (1/g^2,1/g^2,1/g^2)$.
Speed of a null ray in this metric is determined by taking $ds^2=0$ and vanishing of transverse metric; i.e. $dy=0=dz$. It gives $c_N=\frac{dX}{dT}=cg/f$, where $c_N$ is the speed of null ray which is considered to be the velocity of light in the deformed spacetime (\ref{1.01}). Note that the usual speed of light here is scaled by the the energy dependent factors, appearing in the metric. Just like $c$ is invariant in all frames for the usual Minkowski metric, here we consider that the speed of the null ray $c_N$ is invariant in all frames. This means if in another frame the null ray travels with the speed $c'_N=cg'/f'$ then $c_N=c'_N$ implies
\begin{equation}
{\dfrac{g}{f}}={\dfrac{g'}{f'}}~.
\label{new7}
\end{equation}
This will be needed for our future purpose. Remember the notation $f'$ etc. here. $f'$ is the Rainbow function calculated for the transformed energy in the new frame; i.e. $f'\equiv f(E')$.

   Let S be an observer in an inertial coordinate system with coordinates given by $(c_NT,X,y,z)$.\\Another observer $S'$ is traveling in the direction of the $X$ -axis along the trajectory given by $X=f_1(\lambda)$, $T=f_0(\lambda)$ with an arbitrary velocity which may not be constant (Here we have restricted the motion to be only along the X axis to avoid mathematical complexities but it can be generalised to any direction without any issue). $f_1$ and $f_0$ are some specific functions with their explicit form depending on the actual motion, and $\lambda$ is a parameter that is chosen in such a way that it remains invariant under coordinate transformation.  Next we will try to find a suitable coordinate system attached to moving observer $S'$ with the following procedure. For some event P with inertial coordinates ($c_NT,X$) with respect to $S$,the moving observer $S'$ assigns the coordinates ($c_Nt',x'$). At some event A(at $\tau = \tau_A$),light signal is sent to the event P where $\tau$ is the time measured by the moving observer. On reflection from P, the signal is received back at event B (at $\tau = \tau_B$). The coordinates of the moving observer for the event P can be expressed as \cite{15},
\begin{equation}
t'=\frac{1}{2}(\tau_A+\tau_B); \,\,\,\,\
x'=\frac{1}{2}(\tau_B-\tau_A)c_N
\label{1.03}
\end{equation}
The precise expression of $\tau$ will be given later. Next task is to establish a relation between the two coordinate systems ($c_Nt',x'$) and ($c_NT,X$). The inertial coordinates for the above events are given by,
\begin{equation}
X-X_A=c_N(T-T_A); \,\,\,\,\
X-X_B=c_N(T_B-T)~.
\label{1.05}
\end{equation}
Now from (\ref{1.03}) and (\ref{1.05}) we obtain,
\begin{equation}
\tau_A=t'-\frac{x'}{c_N}; \,\,\,\,\
\tau_B=t'+\frac{x'}{c_N}
\label{1.07}
\end{equation}
and
\begin{eqnarray}
&&X-c_NT=X_A-c_NT_A=f_1(\lambda_A) -c_Nf_0(\lambda_A)~;
\label{1.08}
\\
&&X+c_NT=X_B+c_NT_B=f_1(\lambda_B) +c_Nf_0(\lambda_B)~,
\label{1.09}
\end{eqnarray}
respectively. In the last step, the parametric equations for the observer's trajectory have been used.

   Now we need to identify the invariant quantity $\lambda$. This will be done by finding the {\it rainbow} proper time.  For the moving frame there is no spatial change and so the {\it rainbow} proper time is given by
\begin{equation}
-\frac{c^2dt'^2}{f'^2}\equiv\dfrac{-c^2 d\tau^2}{f'^2}=-\frac{c^2dT^2}{f^2}+\frac{d\vec{x}^2}{g^2}~.
\label{1.02}
\end{equation}
This shows that $\tau$ is the {\it rainbow} proper time, given by 
\begin{equation}
d\tau= \frac{f'}{f\gamma}dT; \,\,\ \gamma = \Big(1-\frac{f^2}{c^2}g_{\alpha\beta}v^{\alpha}v^{\beta}\Big)^{-1/2}~,
\label{1.06}
\end{equation}
where $v^\alpha=dx^\alpha/dT$. Now since in the present analysis, the observer is moving along $X$ axis, the velocity $v^{\alpha}$ will have only $X$ component. Then the expression for $\gamma$ reduces to the following form
\begin{equation}
\gamma = \Big(1-\frac{f^2v^2}{c^2g^2}\Big)^{-1/2}~,
\label{gamma}
\end{equation}
where $v=v^{X}=dX/dT$.
So $\tau$ is not an invariant quantity; rather $\tau/f'$ is invariant under coordinate transformations. Hence $\lambda$ can be chosen to be as the {\it rainbow} proper time divided by energy dependent scale factor $f'$. Replacing $\lambda$ by $\tau/f'$ in  Eq. (\ref{1.08}) and Eq. (\ref{1.09}) we obtain
\begin{eqnarray}
X-c_NT=f_1\Big(\dfrac{\tau_A}{f'}\Big) -c_Nf_0\Big(\dfrac{\tau_A}{f'}\Big)~;
\label{1.10}
\\
X+c_NT=f_1\Big(\dfrac{\tau_B}{f'}\Big) +c_Nf_0\Big(\dfrac{\tau_B}{f'}\Big)~.
\label{1.11}
\end{eqnarray}
Next, substitution of the values of ${\tau_A}$ and ${\tau_B}$ from Eq. (\ref{1.07}) in the above equations yield,
\begin{eqnarray}
X-c_NT=f_1\Big(\dfrac{t'}{f'}-\dfrac{x'}{c_N f'}\Big) -c_Nf_0\Big(\dfrac{t'}{f'}-\dfrac{x'}{c_N f'}\Big)~;
\label{1.12}
\\
X+c_NT=f_1\Big(\dfrac{t'}{f'}+\dfrac{x'}{c_N f'}\Big) +c_Nf_0\Big(\dfrac{t'}{f'}+\dfrac{x'}{c_N f'}\Big)~.
\label{1.13}
\end{eqnarray}
The above two relations are the main base for the next analysis. Using them we shall derive the {\it rainbow} Rindler  transformations for an accelerated frame. Before going into that let us briefly show that the {\it rainbow} Lorentz transformations, which are already known \cite{16}, automatically comes from the above two master equations. This will complete our analysis and also give us a verification of the general relations (\ref{1.12}) and (\ref{1.13}) between the inertial and moving frames. 

\section{\label{Lorentz}Rainbow Lorentz transformations}
   In order to obtain the coordinate transformations among two inertial frames where one of them is traveling along $X$ axis with an uniform velocity, one has to first find the parametric trajectory of the moving frame with respect to the other.  
The trajectory of the observer in the inertial frame is $X=vT$ (which is obtained by integrating $v=dX/dT$ with $v$ is constant). Also, in this case, (\ref{1.06}) can be integrated to find the proper time defined as $\tau =(f'T)/(\gamma f)$. So the trajectory parameterized in terms of rainbow proper time turns out to be,
\begin{equation}
T=f_0\Big(\frac{\tau}{f'}\Big)=\dfrac{\gamma f \tau}{f'}; \,\,\,\ X=f_1\Big(\frac{\tau}{f'}\Big)=v\dfrac{\gamma f \tau}{f'}~.
\label{new1}
\end{equation}
Use of these in (\ref{1.12}) and (\ref{1.13}) yields:
\begin{eqnarray}
&&X-c_NT = \Big(\dfrac{t'}{f'}-\dfrac{x'}{c_N f'}\Big)\gamma vf - \Big(\dfrac{t'}{f'}-\dfrac{x'}{c_N f'}\Big)c_N\gamma f~;
\nonumber
\\
&&X+c_NT = \Big(\dfrac{t'}{f'}+\dfrac{x'}{c_N f'}\Big)\gamma vf +\Big(\dfrac{t'}{f'}+\dfrac{x'}{c_N f'}\Big)c_N\gamma f~.
\label{new2}
\end{eqnarray}
Next, solving the above two equation for $x'$ and $t'$ and making use of the relation (\ref{new7}) we obtain transformations among the time and one space (in the direction of motion) coordinates:
\begin{equation}
x'=\dfrac{\gamma g'}{g}\Big(X-vT\Big); \,\,\,\ t'=\dfrac{\gamma f'}{f}\Big(T-\dfrac{vX}{c_N^2}\Big)=\frac{\gamma f'}{f}\Big(T-\frac{f^2v}{g^2c^2}X\Big)~.
\label{1.14}
\end{equation}
The other two coordinates transform as:
\begin{equation}
\frac{y'}{g'}=\frac{y}{g}; \,\,\,\ \frac{z'}{g'}=\frac{z}{g}~.
\label{1.15}
\end{equation}
The above two sets of equations, Eq. (\ref{1.14}) and Eq. (\ref{1.15}), represents the Lorentz transformation between the two frames in the case of rainbow Minkowski metric. One can easily show that these transformations leave the metric (\ref{1.01}) invariant. The same were also obtained earlier in \cite{16} by a different procedure. Here we obtained from a general set of equations (\ref{1.12}) and (\ref{1.13}) which are valid beyond the constant relative velocity among the frames. The recovery of the known result gives us confidence on these equations and hence in the next section we shall find the transformations for going into an uniformly accelerated frame. Before concluding this section, let us point out the following. The rainbow Minkowski metric (\ref{1.01}) takes the usual form in rescaling of the coordinates as $\bar{T}=T/f$, $\bar{X}=X/g$, $\bar{y}=y/g$ and $\bar{z}=z/g$. Then in the bar coordinates the Lorentz transformations are already known. These will be
\begin{eqnarray}
&&\bar{x}' = \bar{\gamma}\Big(\bar{X}-\bar{v}\bar{T}\Big); \,\,\,\ \bar{t}' = \bar{\gamma}\Big(\bar{T}-\frac{\bar{v}\bar{X}}{c^2}\Big);
\nonumber
\\
&&\bar{y}'=\bar{y}; \,\,\,\ \bar{z}'=\bar{z}~,
\label{Lusual}
\end{eqnarray}
where $\bar{\gamma}=(1-\bar{v}^2/{c^2})^{-1/2}$ with $\bar{v}=d\bar{X}/d\bar{T} = (f/g)v$. One can check that these are identical to our derived ones (See Eq. (\ref{1.14}) and Eq. (\ref{1.15})) when we write them in unbar coordinates with $\bar{t'}=t'/f'$, $\bar{x'}=x'/g'$. This again strengths our believe on the results (\ref{1.12}) and (\ref{1.13}) which will be base of the derivation of rainbow Rindler transformations in the next section.

\section{\label{Rindler}Rainbow Rindler transformations and accelerated frame} 
   It may be possible to find the form of the Rindler-Rainbow transformations and metric by just using the scaling argument, like what we noted in the Lorentz transformations case. But one can check that it is very difficult to identify various rainbow quantities in terms of the corresponding usual situation (i.e. in the limit of rainbow parameters goes to unity). Since such a detailed analysis and discussion is lacking in literature we want to fill this gap. We shall see that the ultimate results will give a clear picture in terms of the meaning of different quantities. To do this the general expressions (\ref{1.12}) and (\ref{1.13}) will be used.

    Like the earlier one, we need to find the parametric equations for the trajectory of an uniformly accelerated observer with respect to the inertial frame. These will be obtained, in this case, by knowing the four momentum; i.e. the energy and space momentum of the moving frame with respect to the inertial observer. We shall evaluate them below. 
    
    Here we need to identify the energy and the linear momentum of the moving frame when it is moving on a background (\ref{1.01}) such that they satisfy the dispersion relation (\ref{mdr}). For that let us use the definition of four momentum $p^a=mu^a=m(dx^a/d\tau')$ where $d\tau'$ is the invariant proper time element corresponding to metric (\ref{1.01}). As it was discussed earlier, it is given by $d\tau' = d\tau/f' = dT/f\gamma$. Hence the space component of momentum is given by
\begin{equation}
p^\lambda=mf\gamma\frac{dx^\lambda}{dT} = m{\gamma}fv^{\lambda}~; \,\,\,\ p_\lambda=g_{\lambda\alpha}p^{\alpha}= \frac{m{\gamma}fv_{\lambda}}{g^2}~;
\label{1.19}
\end{equation}
The time component of four momentum is related to the energy as $E/c=-p_a\xi^a$ where $\xi^a$ is the timelike Killing vector for the metric (\ref{1.01}) and chosen as $\xi^a = (1,0,0,0)$. Therefore, it is given by
\begin{equation}
E=-cp_0=-cg_{00}p^0 = \frac{c}{f^2}mf\gamma\frac{d(cT)}{dT}=\dfrac{mc^2\gamma}{f}~.
\label{1.20}
\end{equation}
One can check that the above identification is consistent with the dispersion relation (\ref{mdr}). In that sense it must be argued that the above values for energy and linear momentum are correct.

For the future purpose, we define the magnitude of the space part of the acceleration, linear momentum and velocity of the moving observer as,
\begin{equation}
a = \sqrt{g_{\alpha \beta}a^\alpha a^{\beta}}; \,\,\,\
p = \sqrt{g_{\alpha \beta}p^\alpha p^{\beta}}; \,\,\,\
V = \sqrt{g_{\alpha \beta}v^\alpha v^{\beta}}~.
\label{apv}
\end{equation}
Then using (\ref{1.19}) the linear momentum $p$ can be expressed as
$p = \sqrt{g^{\alpha \beta} m^2 {\gamma}^2 f^2 v_{\alpha} v_{\beta}} = m{\gamma}fV$.
Therefore, the ratio between this and the energy (\ref{1.20}) is $p/E =(f^2 V)/c^2$ from which we find the magnitude of velocity in terms of $p$ and $E$ to be as 
\begin{equation}
V = \frac{pc^2}{Ef^2}~.
\label{new4}
\end{equation}
Now, consider that the motion of the accelerating observer is restricted along the $X$ axis.
Then, the velocity ($V$) reduces to $V = \sqrt{g_{XX} v^{X} v^{X}} = v^X/g = v/g$ while the linear momentum turns out to be $p = p^X/g$. Therefore (\ref{new4}) can be written as
\begin{equation}
v^X =v= \dfrac{gpc^2}{Ef^2}~.
\label{1.22}
\end{equation}

   Let us now concentrate on the equation of motion of the accelerating observer. This will be given by the rate of change of momentum with respect to invariant proper time equals to mass times acceleration. Here the component of linear momentum is $p^{\alpha}$ while the invariant proper time with respect to the inertial frame is $T/f$ (obtained by taking $dx^{\alpha}=0$ in (\ref{1.01})). The effective mass of the particle can be determined by the modified dispersion relation in the limit of $p^{\alpha}=0$.
Putting $p^{\alpha}=0$ in (\ref{mdr}), we have the expression of energy as $E=(mc^2)/f$ which on comparison with $E=Mc^2$ gives us $M=m/f$, which is the effective mass for the present case. Hence, the equation of motion of the moving frame is,
\begin{equation}
\dfrac{dp^{\alpha}}{d({T}/{f})} = \dfrac{ma^{\alpha}}{f}~.
\label{new5}
\end{equation}
As the acceleration is uniform, the above can be integrated to find the linear momentum: $p^{\alpha} = (ma^\alpha T)/f^2$.
For motion along X-axis ,the only non-vanishing component is 
$p^X = (ma^X T)/f^2$. Next using $p=p^X/g$ and $a=a^X/g$ we find $p = (maT)/f^2$. Now,the expression of the acceleration $a$ in terms of that for the usual Minskowki metric $A=(\eta^{\alpha\beta}a_\alpha a_\beta)^{1/2}$, is 
$a = (g_{\alpha\beta} a^\alpha a^\beta)^{1/2}=(g^2 \eta^{\alpha\beta} a_\alpha a_\beta)^{1/2}= gA$. Hence the linear momentum takes the form:
\begin{equation}
p = \dfrac{mATg}{f^2}~,
\label{1.23}
\end{equation}
where A is the acceleration of the moving frame defined in original Minkowski spacetime. 

  It must be pointed out that in this analysis the metric (\ref{1.01}) has been taken as a classical background which incorporates the modification due to the energy of the moving particle on the background. But when we are considering the accelerated frame, it is assumed that this frame is not modifying the background due to its motion. If there is any correction by the presence of particle, that is already presented in $f$ and $g$ of modified Minkowsi metric (\ref{1.01}). This is fine as we are interested to find the metric of moving frame which already being modified and hence there is no need to modify again due to energy of the accelerated frame. Of course, the energy and momentum of the frame must be such that it satisfies the same form of modified dispersion relation (\ref{mdr}).  Therefore $f$ and $g$, which are functions of energy of the existing particle on the Minkowski background, will act as constant variable with respect to the accelerated frame energy. This has been used in the above and will be used later also. The same trick is always followed in studying the black hole thermodynamics in rainbow gravity. In this way we shall find the Rindler metric modified by the existing particles' energy residing on the spacetime.

    Next substituting values of $E$ and $p$ from (\ref{mdr}) and (\ref{1.23}) in  (\ref{1.22}), we find the differential form of the trajectory of the accelerated frame as:
\begin{equation}
\dfrac{dX}{dT} = \dfrac{g^2T}{f^2(\frac{f^2}{A^2}+\frac{T^2g^2}{c^2f^2})^{\frac{1}{2}}}~.
\label{new6}
\end{equation}
Integrating the above, one finds the equation of the trajectory:
\begin{equation}
\dfrac{X^2}{g^2}-\dfrac{c^2T^2}{f^2} = \dfrac{c^4f^2}{A^2g^2} \equiv \dfrac{c^4}{\tilde{A}^2}~,
\label{1.24}
\end{equation}
where, $ \tilde{A} = {Ag}/f$.

  Now, our aim is to find the parametric equations of the trajectory of the accelerated frame. For that, start with Eq. (\ref{1.06}), which for $X$ direction of motion reduces to,
 \begin{equation}
 d\tau = dT \frac{f'}{f}\Big(1-\frac{f^2v^2}{c^2g^2}\Big)^{1/2} = dT\frac{f'}{f}\Big(1 - \frac{p^2c^2}{E^2f^2}\Big)^{1/2}~,
 \label{4.01}
 \end{equation}
 where, in the last step the value of $v$ from (\ref{1.22}) has been used. Next, substituting the values of $E$ and $p$ from (\ref{mdr}) and (\ref{1.23})  respectively in the above and then integrating we obtain:
\begin{equation}
\tau = \dfrac{f'}{f} \int \dfrac{dT}{\sqrt{1+\frac{A^2g^2T^2}{c^2f^4}}}
= \dfrac{cff'}{Ag} \sinh^{-1}\Big(\frac{AgT}{cf^2}\Big)~.
\label{4.02}
\end{equation}
Therefore, the parametric equation for $T$ is,
\begin{equation}
T=f_0\Big(\frac{\tau}{f'}\Big) = \dfrac{cf^2}{Ag}\sinh\Big(\frac{Ag\tau}{cff'}\Big)~.
\label{1.25}
\end{equation}
The other equation for $X$ is calculated by substituting the above in (\ref{1.24}). This leads to, 
\begin{equation}
X = f_1\Big(\frac{\tau}{f'}\Big) = \dfrac{c^2f}{A}\cosh\Big(\frac{Ag\tau}{cff'}\Big)~.
\label{1.26}
\end{equation}

  We are now in a position to find the coordinate transformations from inertial frame to accelerated frame. This will be done, like the Lorentz transformations, by using the parametric equations (\ref{1.25}) and (\ref{1.26}) in the general equations (\ref{1.12}) and (\ref{1.13}). This yields,
\begin{equation}
X-c_NT = \dfrac{c^2f}{A}~\exp\Big[-\frac{Ag}{cff'}\Big(t'-\frac{x'}{c_N}\Big)\Big]~;
\label{4.03}
\end{equation}
and
\begin{equation}
X+c_NT= \dfrac{c^2f}{A} ~\exp\Big[\frac{Ag}{cff'}\Big(t'+\frac{x'}{c_N}\Big)\Big]~.
\label{4.04}
\end{equation}
Adding and subtracting above two equations and using $c_N=(cg)/f$, we get:
\begin{eqnarray}
X=\dfrac{c^2f}{A}~e^{\dfrac{Ax'}{c^2 f'}}\cosh\Big({\frac{Ag}{cff'}t'}\Big)~;
\label{1.27}
\\
T=\dfrac{cf^2}{Ag}~e^{\dfrac{Ax'}{c^2 f'}} \sinh\Big({\frac{Ag}{cff'}t'}\Big)~.
\label{1.28}
\end{eqnarray}
This gives the transformation between the inertial coordinate system and that of a uniformly accelerated observer. The other two coordinates transform in the similar way as given  by Eq. (\ref{1.15}).
The rainbow Rindler metric is therefore of the form,
\begin{equation}
ds^2 =e^{\dfrac{2Ax'}{c^2 f'}}\Big(\dfrac{-c^2dt'^2}{f'^2}+\dfrac{dx'^2}{g'^2}\Big)+\frac{dy'^2}{g'^2}+\frac{dz'^2}{g'^2}~.
\label{1.29}
\end{equation}
Now going to the energy dependent coordinates $t = t'/f'$, $x = x'/g'$, $\tilde{y}=y'/g'$ and $\tilde{z}=z'/g'$ we express the Rindler metric as,
\begin{equation}
ds^2 = e^{\dfrac{2Ag'x}{c^2 f'}}\Big(-c^2dt^2+dx^2\Big)+dL_{\perp}^2=e^{\dfrac{2Agx}{c^2 f}}\Big(-c^2dt^2+dx^2\Big)+dL_{\perp}^2~,
\label{1.30}
\end{equation}
where $dL_{\perp}^2=d\tilde{y}^2+d\tilde{z}^2$ is the transverse metric and in the last step (\ref{new7}) has been used.

  Let us now mention, why the above detailed analysis is important. Note that, in this process the correct expression for the effective acceleration has been identified. We found that it is given by $\tilde{A}=(gA)/f$ where $A$ is the usual acceleration of the moving frame. Below, we shall show that the Unruh temperature will be given by $\tilde{A}/(2\pi)$ (with $c=1$) which is compatible with horizon temperature of the black hole in rainbow gravity (see \cite{18} for the value of Hawking temperature in the case of Rainbow-Schwarzschild spacetime). One may check that all these main explicit expressions (Eqs. (\ref{1.27}), (\ref{1.28}) and (\ref{1.29}) or (\ref{1.30})) are very hard to come by using simple scaling analysis. Therefore, we feel that the above discussion is necessary in this paradigm.  

\section{\label{Unruh}Unruh effect in rainbow Rindler metric}

    Here,we will determine the Unruh temperature in the rainbow Rindler metric by three well known approaches: Fourier coefficient method, Bogolyubov coefficients and Detector response method. We will discuss Fourier coefficient method in detail  and provide only the important conclusions from the other two approaches. 

\subsection{\label{Fourier}Fourier coefficient method}
The Unruh temperature is found below by studying the power spectrum, observed by an accelerated observer in rainbow frame. The solution, corresponding to the modified dispersion relation (\ref{mdr}), under the background (\ref{1.01}) is the plane wave solution of the form: $\phi = \exp[-i{\Omega}({{T}/{f}}-{{X}/{g}})]$ {\footnote{From now we shall take the unit such that $c=1$.}}. Now to see how an accelerating observer will measure it in its own frame, one has to write ($T,X$) in terms of the rainbow proper time, given by the transformations (\ref{1.25}) and (\ref{1.26}). In this frame, $\phi$ takes the form 
\begin{equation}
\phi = \exp~ \bigg [i\bigg( {\dfrac {\Omega f}{Ag}}~\exp{\dfrac{-Ag\tau}{ff'}}\bigg)\bigg]
\label{1.31}
\end{equation}
The Fourier transform of the above solution is given by,
\begin{equation}
\phi\bigg({\dfrac{\tau}{f'}}\bigg)~=~ \int^{+\infty}_{-\infty}\frac{d\nu}{2\pi}\  f(\nu)e^{{-i\nu\tau}/{f'}}
\end{equation}
\begin{equation} 
f(\nu)~=~ \int^{+\infty}_{-\infty} \frac{d\tau}{f'}\phi\bigg(\dfrac{\tau}{f'}\bigg)e^{{i\nu\tau}/{f'}}~. 
\end{equation}
where $f(\nu)$ is the Fourier transform of $\phi({\tau}/{f'})$ with respect to $({\tau}/{f'})$.
The power spectrum of this wave can be determined by $|f(\nu)|^2$. 
In the above, the integration variable in the second one has been taken as $d\tau /f'$ instead of just $d\tau$. This is because we observed earlier that the latter is not an invariant quantity under rainbow transformation.
Evaluating the above Fourier transform one gets 
\begin{equation}
f(\nu)~=~\bigg(\dfrac{1}{\tilde{A}}\bigg)\bigg(\dfrac{\Omega}{\tilde{A}}\bigg)^{{i\nu}/{\tilde{A}}}~\Gamma\big({-i\nu}/{\tilde{A}}\big)~e^{{\pi\nu}/{2\tilde{A}}}~,
\end{equation}
where $\tilde{A}={Ag}/{f}$. 
Obtaining the modulus $|f(\nu)|^2 $ ,we get the remarkable result that the power, per logarithmic band in frequency , at negative frequencies is a Planckian :
\begin{equation}
\nu|f(-\nu)|^2~=~ \dfrac{2\pi}{\tilde{A}(e^\frac{2\pi\nu}{\tilde{A}} -1)}~.
\end{equation}
From the above one identifies the temperature as
\begin{equation}
\tilde{T}_0 =\dfrac{\tilde{A}}{2\pi} = (T_0g)/f~
\label{temp}
\end{equation}
where $T_0=A/2\pi$ is the usual expression for temperature.

   One can notice that the above expression for temperature is consistent with the time-like Killing vector corresponding to the metric (\ref{1.30}).  It is well known that the expression for the surface gravity corresponding to the metric $ds^2=F(x)(-dt^2+dx^2)+d\Omega^2$ is given by $\kappa = F'(x_0)/2$ where $x_0$ is the location of the horizon. This is obvious from the expression $\kappa= -1/2(\nabla^b\xi^a)(\nabla_b\xi_a)$ where $\xi^a$ is the time-like Killing vector for this metric. Now (\ref{1.30}) has $F(x) = \exp[(2Agx)/f]$ and hence $\kappa=(Ag)/f$. Therefore the temperature is given by (\ref{temp}) and hence it is compatible with the time-like Killing vector of the metric.

\subsection{\label{Bogo}Bogolyubov coefficients}
The equation of motion for the scalar field $\Phi$ is determined by solving the equation,
\begin{equation}
\Box\Phi = \dfrac{1}{\sqrt{-g}}\partial_\mu\bigg(\sqrt{-g}~g^{\mu\nu}\partial_\nu \bigg)\Phi=0~.
\label{1.32}
\end{equation}
The normalized solution to this equation are plane waves i.e
\begin{equation}
u_{k}\bigg(T,X\bigg)= \dfrac{1}{\sqrt{4\pi\omega}}\exp\Big[-i\bigg(\dfrac{\omega T}{f}- \dfrac{kX}{g}\bigg)\Big]
\label{1.33}
\end{equation}
where $\omega=|k|$ and $k$ can take values continuously in the range $-\infty$ and $\infty$.
We find the modes (\ref{1.33}) in such a way that they satisfy the following orthnormality relations \cite{17}:
\begin{equation}
(u_k ,u_{k'})=\delta_D (k - k ')~~~~~~~~(u^*_k ,u^*_{k '})=-\delta_D (k - k ')~~~~~~~~(u_k ,u^*_{k '})=0
\label{1.34}
\end{equation}
Similarly the solutions to equation (\ref{1.32}) for the rainbow Rindler metric given by equation (\ref{1.29}) are also found to be plane waves which follow the same form of orthonormalization relation with $u_k$ and $u_k'$ being replaced by $v_l$ and $v_l'$ respectively. These are given by
\begin{equation}
v_{l}(t',x')= \dfrac{1}{\sqrt{4\pi \nu}}\exp-\-i\bigg(\dfrac{\nu t'}{f'}-\dfrac{lx'}{g'}\bigg)
\end{equation}
where $\nu=|l|$ and $l$ can take values continuously between $-\infty$ and $\infty$.
The Bogolyubov coefficients $\alpha(l,k)$ and $\beta(l,k)$ are given by the expression \cite{17},
\begin{equation}
\alpha(l,k)~=~ (v_l ,u_k)~~~~~~~~~ \beta(l,k)~=~-(v_l,{u^*}_k)
\end{equation}
Carrying out these integrals we find the explicit forms of the Bogolyubov coefficients: 
\begin{equation}
\alpha(l,k)~=~\bigg(\dfrac{\tilde{A}^{-1}}{4\pi k\sqrt{\omega \nu}}\bigg)(\omega l + k \nu)(k \tilde{A}^{-1})^{-\-i lg^{-1}}\Gamma (il\tilde{A}^{-1})e^{\pi l/{2\tilde{A}}}
\end{equation}
\begin{equation}
\beta(l,k)~=~-\alpha(l,k)e^{-\pi l/{\tilde{A}}}~.
\end{equation}
Since $\alpha(l,k)$ and $\beta(l,k)$ are constant complex coefficients obeying the condition:
\begin{equation}
|\alpha|^2 - |\beta|^2~=~1~.
\end{equation}
Solving it for $|\beta|^2$ we get,
\begin{equation}
|\beta|^2~=~\dfrac{1}{e^{2\pi l/\tilde{A}} - 1}~,
\end{equation}
which is the emission spectrum and is again Planckian. The temperature again turns out to be that given by (\ref{temp}).

\subsection{\label{Detector}Detector response}
For a detector in its ground state,the probability of transition to one of it's excited state, to the lowest order in the perturbation theory, is given by{\cite{15}},
\begin{equation}
\mathcal{P}(E)~=~|\mathcal{M}|^2~ \int^{\infty}_{-\infty}(d{\tau/f'})~\int^{\infty}_{-\infty}(d{\tau'/f'})~e^{-iE(\tau-\tau')/f'}~G[x(\tau/f'),x(\tau'/f')]~,
\label{1.35}
\end{equation}
where $ |\mathcal{M}|^2\equiv~|\langle E_1|\mu(0)|E_0 \rangle|^2~,E~\equiv~E_1-E_0$, $E_1$ and $E_0$ are the energy of the excited and ground states of the detector,and G is the Wightman function given by \cite{17},
\begin{equation}
G[x(\tau/f'),x(\tau'/f')]~=~\langle 0|\Phi[x(\tau/f')]\Phi[x(\tau'/f')]|0\rangle~.
\end{equation}
The explicit form of the Wightman function for the rainbow Rindler metric is given by,
\begin{equation}
G=-\dfrac{1}{4\pi}\ln\Big[\dfrac{4}{\tilde{A}^2}\sinh^2\Big(\dfrac{\tilde{A}(\Delta{\tau}/f' - i\epsilon)}{2}\Big) \Big]
\label{1.36}
\end{equation}
In the equation (\ref{1.35}),the important part is the response function per unit time that gives the rate excitation,
\begin{equation}
\mathcal{R}(E)~=~ \int^{\infty}_{-\infty}~ e^{-iE\Delta \tau /f'}G(\Delta \tau /f')d(\Delta \tau /f')
\end{equation}
Using the value of the Wightman function($G$) from(\ref{1.36}) in the above relation and performing the contour integral,the rate of excitation comes out to be, 
\begin{equation}
\mathcal{R}(E)~ \propto ~\Big(\dfrac{E}{2\pi}\Big)\Big(\dfrac{1}{e^{2\pi E/{\tilde{A}}}-1}\Big)~.
\end{equation}
This is precisely what one would have found in the case of a detector immersed in a thermal bath with temperature  $\tilde{A}/2\pi$ ~.
\section{\label{Conclusions}Discussions}
In this paper, we have used the functional forms of transformations from rainbow Minkowski spacetime to any moving frame and hence used them to verify the already existing Rainbow Lorentz transformations. Keeping in view of the importance of Rindler frame in studying relativistic (GR) features and black hole thermodynamics, Rainbow-Rindler transformations were obtained.
In this process, all the parameters are properly and explicitly identified. The effective acceleration for the deformed spacetime was obtained as $\tilde {A}=(gA)/f $, where, A is the usual acceleration (i.e. in the limit of rainbow parameters goes to unity) of the moving frame. 
Finally, using the transformations the Unruh temperature has also been evaluated as $\tilde{A}/2\pi$ through three distinct approaches: Fourier coefficient, Bogolyubov coefficients and Detector response methods.
This form of Unruh temperature exactly agrees with Horizon temperature of the black hole in Rainbow gravity. 

One of the importance of this analysis is as follows. The obtained expression for rainbow-Unruh temperature can be used to find that for the black hole horizon. There are two ways to look at this point of view. 
\vskip 1mm
\noindent
(i) We have stated earlier that the near horizon metric of a black hole is effectively in Rindler form. Therefore the temperature obtained here (i.e. Unruh expression) exactly mimics that of the black hole horizon in rainbow context. Incidentally, this form of temperature for black holes in rainbow gravity was taken earlier as an input without any clear justification. Our present analysis strongly supports that. 
\vskip 1mm
\noindent 
(ii) The same conclusion can also be drawn by the concept of global embedding of black hole spacetime in the Minskowski space (GEMS) approach (for original works see, \cite{Deser:1998bb,Deser:1998xb}; one may also see the ref. \cite{Banerjee:2010ma} for more effective way of looking at the approach) which helps to understand the Unruh and Hawking effects in a unified point of view. For the usual cases, one first finds the embedding flat spacetime. Then it has been observed that a particular observer on this will follow the uniformly accelerated observer trajectory. Therefore the identification of horizon temperature by the concept of Unruh effect naturally enters. For rainbow black hole metric, same path can also be taken. Of course, this requires more investigations as the embedding flat space is not known in rainbow gravity.

In any case one has to identify the correct form of Unruh expression. In a nut shell, it is obvious that the Rindler form of metric plays several important roles in revealing thermodynamics of gravity. Therefore it is quite natural to obtain the same in rainbow gravity. This is precisely obtained here.   
Also, since the study of accelerated frame provides major insights into knowing gravity with more depth, we believe the above result will provide a right pathway to explore new dimensions of Rainbow Gravity.

\vskip 9mm
\noindent
{\bf{Acknowledgements}}\\
\noindent
The research of one of the authors (B.R.M.) is supported by a START-UP RESEARCH GRANT (No. SG/PHY/P/BRM/01) from Indian Institute of Technology Guwahati, India.

\end{document}